\documentclass[pra,twocolumn,superscriptaddress,showpacs,nofootinbibfloatfix,amsmath,amsfonts,amssymb]{revtex4-1}%
\usepackage{amsmath,amsfonts,amssymb,color}
\usepackage{amsthm}
\usepackage{leftidx}
\usepackage{graphicx}
\usepackage{xcolor}
\usepackage{dcolumn}
\usepackage{bm}
\usepackage{epstopdf}
\usepackage{epsfig}
\usepackage{environ}
\usepackage{pdfcomment}

\usepackage{float}
\usepackage[T1]{fontenc}
\usepackage[latin9]{inputenc}
\usepackage{setspace}
\usepackage{esint}

\begin{document}
	
\preprint{APS/123-QED}

\title{Driving-induced multiple ${\cal PT}$-symmetry breaking transitions and reentrant localization
	transitions in non-Hermitian Floquet quasicrystals}

\author{Longwen Zhou}
\email{zhoulw13@u.nus.edu}
\affiliation{%
	College of Physics and Optoelectronic Engineering, Ocean University of China, Qingdao, China 266100
}

\author{Wenqian Han}
\affiliation{%
	College of Physics and Optoelectronic Engineering, Ocean University of China, Qingdao, China 266100
}

\date{\today}

\begin{abstract}
The cooperation between time-periodic driving fields and non-Hermitian
effects could endow systems with distinctive spectral and transport
properties. In this work, we uncover an intriguing class of non-Hermitian
Floquet matter in one-dimensional quasicrystals, which is characterized
by the emergence of multiple driving-induced ${\cal PT}$-symmetry
breaking/restoring transitions, mobility edges, and reentrant localization transitions.
These findings are demonstrated by investigating the spectra, level
statistics, inverse participation ratios and wavepacket dynamics of
a periodically quenched nonreciprocal Harper model. Our results not
only unveil the richness of localization phenomena in driven non-Hermitian
quasicrystals, but also highlight the advantage of Floquet approach
in generating unique types of nonequilibrium phases in open systems.
\end{abstract}

\maketitle

\section{Introduction\label{sec:Int}}
Periodic driving fields could create dynamical states of matter that
are absent in equilibrium settings, with Floquet topological phases
\cite{FTPRev1,FTPRev2,FTPRev3,FTPRev4,FTPRev5} and discrete time
crystals \cite{DTC1,DTC2,DTC3,DTC4,DTC5} being two representative
examples. The discovery of these unique phases not only extends the
classification of quantum matter to nonequilibrium situations \cite{TPClass1,TPClass2,TPClass3,TPClass4,TPClass5,TPClass6,TPClass7,TPClass8,TPClass9},
but also leads to breakthroughs in the experimental characterization
of quantum dynamics in complex systems \cite{FPExp1,FPExp2,FPExp3,FPExp4,FPExp5,FPExp6,FPExp7,FPExp8,FPExp9,FPExp10,FPExp11,FPExp12,FPExp13,FPExp14}.

In recent years, possible new phases that could emerge due to the
interplay between Floquet drivings and non-Hermitian effects are considered.
Theoretical progresses have been made to the discovery of \emph{non-Hermitian
	Floquet} topological insulators \cite{NHFTI1,NHFTI2,NHFTI3,NHFTI4,NHFTI5,NHFTI6,NHFTI7,NHFTI8,NHFTI9,NHFTI10,NHFTI11},
second-order topological phases \cite{NHFSOTP1,NHFSOTP2}, topological
superconductors \cite{NHFTSC1,NHFTSC2} and semimetals \cite{NHFSM1,NHFSM2,NHFSM3,NHFSM4}.
Intriguing features such as non-Hermiticity induced Floquet topological
edge states \cite{NHFSOTP1} and their coexistence with Floquet non-Hermitian
skin effects \cite{NHFTI8} have also been uncovered. In experiments,
setups like cold atoms \cite{NHFCdAt1} and photonics \cite{NHFPhot1,NHFPhot2,NHFPhot3,NHFPhot4,NHFPhot5}
have been demonstrated as efficient platforms to explore phases and
transitions in driven non-Hermitian systems. Beyond the clean limit,
high-frequency driving fields have been found as a flexible knob to
control the spectral, localization and topological transitions in
non-Hermitian quasicrystals \cite{NHFQC1}. However, the main power
of Floquet engineering in generating unique nonequilibrium phases,
which usually displays itself at resonant driving frequencies and
amplitudes has yet to be unveiled for non-Hermitian disordered systems.

A non-Hermitian static system described by a ${\cal PT}$-symmetric Hamiltonian $H$ can undergo a ${\cal PT}$ transition when the energy spectrum of $H$ switches from real to complex (or vice versa) with the change of system parameters. On one side of the transition point, all the eigenvalues of $H$ are real and the system is in a ${\cal PT}$ invariant (unbroken) phase with all eigenstates being ${\cal PT}$-symmetric. On the other side of the transition point, certain eigenenergies have nonvanishing imaginary parts and the system is in a ${\cal PT}$ broken phase, in which certain eigenstates are not ${\cal PT}$-symmetric. Comparatively, a Floquet system can undergo a ${\cal PT}$ transition when the quasienergy spectrum of the system's Floquet operator $U$ goes from real to complex (or vice versa) with the change of system parameters. On one side of the transition, all the quasienergies $E$ of $U$ obtained from the eigenvalue equation $U|\psi\rangle=e^{-iE}|\psi\rangle$ are real, and the system is in a ${\cal PT}$ invariant (unbroken) phase with all Floquet eigenstates $|\psi\rangle$ being ${\cal PT}$-symmetric. On the other side of the transition, certain quasienergies of $U$ have nonvanishing imaginary parts and the system is in a ${\cal PT}$ broken phase with certain non-${\cal PT}$-symmetric Floquet eigenstates. The ${\cal PT}$ transition now shows up in the quasienergy spectrum of a Floquet operator, instead of the energy spectrum of a Hamiltonian.

\begin{figure}
	\begin{centering}
		\includegraphics[scale=0.39]{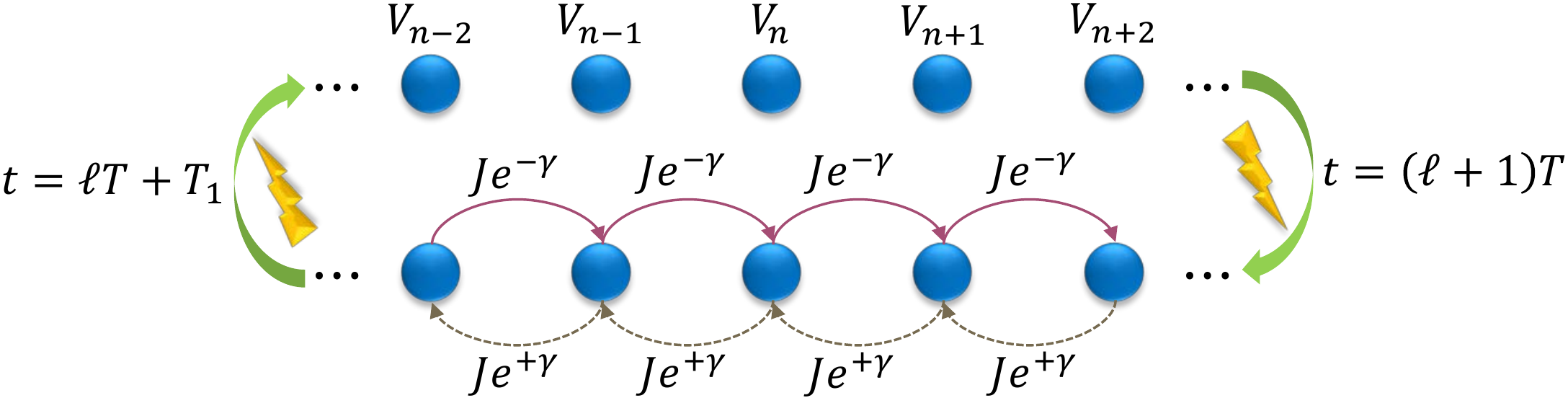}
		\par\end{centering}
	\caption{Schematic plot of the periodically quenched nonreciprocal Harper model.
		In the upper chain, $V_{n}=V\cos(2\pi\alpha n)$ is the potential
		on the $n$th lattice site, which is spatially quasiperiodic for an
		irrational $\alpha$. $Je^{\mp\gamma}$ couple nearest sites of the
		lower chain in an asymmetric manner when $\gamma\protect\neq0$. Within
		each driving period $T=T_{1}+T_{2}$, the configuration of the system
		is switched between the upper and lower chains following the quench
		protocol described by Eq.~(\ref{eq:Ht}). $\ell\in\mathbb{Z}$ counts
		the number of driving periods.\label{fig:Sketch}}
\end{figure}

In the present work, we show that going beyond high-frequency modulations,
periodic driving fields could endow non-Hermitian quasicrystals with
multiple ${\cal PT}$ spectral transitions, reentrant localization
transitions, richer phase diagrams and unique dynamical properties.
These facts are demonstrated explicitly in a periodically quenched
nonreciprocal Harper model (NRHM), as introduced in Sec.~\ref{sec:Mod}.
In Sec.~\ref{sec:Met}, we provide a theoretical framework to study
the spectrum and transport properties of one-dimensional (1D) non-Hermitian
Floquet quasicrystals. These methods are then applied to characterize
the extended, critical, localized phases and the sequence of ${\cal PT}$
and localization transitions among them in the periodically quenched
NRHM in Sec.~\ref{sec:Res}. Wavepacket dynamics is further employed
as a probe to signify non-Hermitian Floquet quasicrystals with
different transport nature. 
The physical origins of rich non-Hermitian Floquet quasicrystal phases and transitions are discussed in Sec.~\ref{Dis}.
We summarize our results and discuss potential
future directions in Sec.~\ref{sec:Sum}.
Some other numerical details are presented in the Appendixes \ref{app1} and \ref{app2}.

\section{Model\label{sec:Mod}}
Non-Hermitian quasicrystal (NHQC) forms a class of matter with unique 
spectral, transport and topological nature due to the interplay between
lattice quasiperiodicity and non-Hermitian effects~\cite{NHQC1,NHQC2,NHQC3,NHQC4,NHQC5,NHQC6,NHQC7,NHQC8,NHQC9,NHQC10,NHQC11,NHQC12,NHQC13,NHQC14,NHQC15}.
In this section, the setting of a periodically quenched NRHM is introduced,
which is originated from a prototypical model in the study of NHQC.

The Hamiltonian of the static NRHM takes the form $\hat{H}=\hat{K}+\hat{V}$,
where in the lattice basis $\{|n\rangle\}$ we have
\begin{alignat}{1}
	\hat{K}& = J\sum_{n}(e^{\gamma}|n\rangle\langle n+1|+e^{-\gamma}|n+1\rangle\langle n|),\label{eq:K}\\
	\hat{V}& = V\sum_{n}\cos(2\pi\alpha n)|n\rangle\langle n|.\label{eq:V}
\end{alignat}
Here $n\in\mathbb{Z}$ is the lattice index. $Je^{\pm\gamma}$ are
the nearest-neighbor hopping amplitudes. $V$ is the amplitude of
the onsite potential $V_{n}=V\cos(2\pi\alpha n)$. We assume $J,V>0$
without loss of generality. When $\gamma\neq0$, the hoppings from
the right to left and from the left to right neighboring sites become asymmetric,
yielding a non-Hermitian Hamiltonian~($\hat{H}\neq\hat{H}^{\dagger}$).
Meanwhile, an irrational $\alpha$ makes the onsite potential quasiperiodic,
leading to a 1D NHQC \cite{NRHM1,NRHM2}. 
The spectrum of $\hat{H}$, obtained by solving the eigenvalue equation $\hat{H}|\psi\rangle=E|\psi\rangle$	can be real in certain parameter regions due to the ${\cal PT}$ symmetry. To see this, we first take the periodic boundary condition (PBC) for $\hat{H}$ and the rational approximation $\alpha\simeq p/q$ for $\alpha$. For example, if we set $\alpha=(\sqrt{5}-1)/2$, $p$ and $q$ can take two adjacent numbers in the Fibonacci sequence (with $p<q$) to form the rational approximation $\alpha\simeq p/q$. Applying the discrete Fourier transformation $|n\rangle=\frac{1}{\sqrt{L}}\sum^L_{l=1}|l\rangle e^{i2\pi\alpha l n}$ to ${\hat H}$ \cite{NHQC4}, we obtain its representation in momentum space as ${\hat H}'=\frac{V}{2}\sum_l (|l\rangle\langle l+1|+{\rm H.c.})+2J\sum_l \cos(2\pi\alpha l+i\gamma)|l\rangle\langle l|$. Note that the quasimomentum here is $2\pi\alpha l$ and the length of lattice is $L=q$. We can now identify in momentum representation the ${\cal P}$ symmetry of the system as ${\cal P}=\sum_l |-l\rangle\langle l|$ and the ${\cal T}$ symmetry as the complex conjugate ${\cal K}$. The off-diagonal part of ${\hat H}'$ is Hermitian and clearly symmetric under the combined ${\cal PT}$ operation. The diagonal elements $W_n=2J\cos(2\pi\alpha l+i\gamma)$ satisfy $W_n=W^*_{-n}$, which guarantees the invariance of the diagonal part of ${\hat H}'$ under the ${\cal PT}$ operation. Putting together, we have $[{\cal PT},{\hat H}']=0$. The momentum-space representation of the system Hamiltonian ${\hat H}$ is thus symmetric under ${\cal PT}=\sum_l|-l\rangle\langle l|{\cal K}$. Going back to the lattice representation, we may express the ${\cal PT}$ symmetry operator as ${\cal PT}=(\sum_n |-n\rangle\langle n|)({\cal K}U_{\rm FT})$, where the Fourier transformation $U_{\rm FT}$ has the matrix elements $(U_{\rm FT})_{n,l}=\frac{1}{\sqrt{L}}e^{-i2\pi\alpha nl}$. In fact, as ${\hat H}$ and ${\hat H}'$ are only differ by a unitary Fourier transformation that is independent of the system parameters $(J,V,\gamma)$, their spectra are identical under the PBC. The ${\cal PT}$-breaking transitions in the spectrum of ${\hat H}'$ are thus coincident with the real-to-complex spectrum transitions in ${\hat H}$. For example, with the increase of $\gamma$, the non-Hermitian effects become stronger, and complex eigenenergies of $\hat{H}$ start to appear after the system undergoes the ${\cal PT}$-breaking transition. Under the PBC, such a transition was found to occur at $\gamma=\gamma_{c}=-\ln(2J/V)$ for $V>2J>0$~\cite{NRHM1}.
Interestingly, when $\gamma$ changes from $\gamma_{c}+0^{+}$ to
$\gamma_{c}-0^{+}$, the energies of all eigenstates change from complex
to real and their spatial profiles switch from extended to 
localized with the common Lyapunov exponent (inverse localization
length) $\lambda=\ln[Ve^{-\gamma}/(2J)]$ \cite{NRHM1}. Therefore,
we obtain a ${\cal PT}$ transition in conjunction with a localization
transition for all eigenstates at a finite amount of hopping asymmetry
$\gamma=\gamma_{c}$ in the NRHM. 
With Floquet periodic drivings, the ${\cal PT}$ transition points in the NRHM can be
flexibly controlled \cite{NHFQC1}. Alternating transitions between extended and localized
phases with distinct topological nature can be further induced via changing the driving field parameters.
However, no mobility edges and intermediate
phases are found in the presence of high-frequency driving fields \cite{NHFQC1}.

In this work, we focus on a periodically quenched variant of the NRHM,
which goes beyond the fast modulation protocol considered in
previous studies~\cite{NHFQC1}. The time-dependent Hamiltonian of the
periodically quenched NRHM takes the form
\begin{equation}
	\hat{H}(t)=\begin{cases}
		\hat{K} & t\in[\ell T,\ell T+T_{1})\\
		\hat{V} & t\in[\ell T+T_{1},\ell T+T_{1}+T_{2})
	\end{cases}.\label{eq:Ht}
\end{equation}
Here $T=T_{1}+T_{2}$ is the driving period. $\ell\in\mathbb{Z}$
counts the number of periods in the evolution. $\hat{K}$ and $\hat{V}$
are given by Eqs.~(\ref{eq:K}) and (\ref{eq:V}), respectively. An
illustration of this time-dependent lattice and the driving protocol
is given in Fig.~\ref{fig:Sketch}. The Floquet operator of the system,
which corresponds to its evolution operator over a complete driving
period reads
\begin{equation}
	\hat{U}=e^{-i\mathsf{V}\sum_{n}\cos(2\pi\alpha n)|n\rangle\langle n|}e^{-i\mathsf{J}\sum_{n}(e^{\gamma}|n\rangle\langle n+1|+e^{-\gamma}|n+1\rangle\langle n|)}.\label{eq:U}
\end{equation}
Here we have introduced the dimensionless parameters $\mathsf{V}=VT_{2}/\hbar$
and $\mathsf{J}=JT_{1}/\hbar$, with $\hbar$ being the Planck constant.
As an advantage of Floquet engineering, the amplitudes of hopping
and onsite potential can be easily and separately tuned by varying
the time durations $T_{1}$ and $T_{2}$ of the piecewise quench protocol,
making it more flexible to control spectral and localization transitions
in the system. Note that such a Floquet operator can also be realized
by making the hopping amplitudes asymmetric in the kicked Harper model
(KHM) \cite{KHM1,KHM2,KHM3}. The later has been found to possess
rich Floquet topological insulating phases in the Hermitian regime
\cite{KHM4,KHM5,KHM6}. We can thus regard the Floquet operator in
Eq.~(\ref{eq:U}) equivalently as describing a nonreciprocal KHM,
or NRKHM in short. Interestingly, Anderson transitions in quasiperiodic
KHM (with irrational $\alpha$) was also investigated in early studies
\cite{KHM1,KHM2,KHM3}. While non-Hermitian effects on the spectral and
localization transitions in the KHM have yet to be revealed. Throughout
this work, we will choose $\alpha$ to be the inverse golden ratio,
i.e., $\alpha=\frac{\sqrt{5}-1}{2}$ in order to endow the lattice
with spatial quasiperiodicity. We will also perform all our numerically
studies of the $\hat{U}$ in Eq.~(\ref{eq:U}) under the PBC, such that
the possible influence non-Hermitian skin effect is absent.

The NRKHM considered here is closely related to experimentally realizable physical systems. Recently, a quasiperiodic KHM was realized by applying apodized Floquet engineering techniques to ultracold $^{84}$Sr atoms \cite{KHM7}. In the Hermitian limit ($\gamma=0$), the Floquet operator of our model is exactly equivalent to the one realized in Ref.~\cite{KHM7}. The direct realization of finite nonreciprocal hopping is challenging for cold atom setups. Recently, techniques to effectively engineer nonreciprocal hopping for cold atoms have been considered in both theory and experiments by introducing atomic losses \cite{NHFCdAt1,NRHP1,NRHP2,NRHP3}. Cold atom systems are thus good candidates to realize our model and probe its physical properties in near-term experiments. On the other hand, the ${\cal PT}$ and localization transitions in a temporally driven (Floquet) dissipative quasicrystal was recently observed \cite{NHFPhot5}. The experiment there is implemented by photonic quantum walks in coupled optical fibre loops. The model realized in Ref.~\cite{NHFPhot5} may be viewed as a dynamical variant of the NRHM, which contains two key elements for the realization of our model, i.e., the piecewise periodic quench and the nonreciprocal hopping. Meanwhile, only a transition between localized and delocalized phases accompained by the breaking of ${\cal PT}$ symmetry is observed in Ref.~\cite{NHFPhot5} due to the short-range nature of the realized hopping amplitudes. By implementing long-range steps to the walker in modified versions of the system in Ref.~\cite{NHFPhot5}, the evolution operator for the hopping part of our model may be realized. Therefore, photonic quantum walks can also serve as a candidate to realize our system and detect the multiple ${\cal PT}$ and reentrant localization transitions there.

\section{Method\label{sec:Met}}
In this section, we outline essential tools that will be employed to characterize
various non-Hermitian Floquet quasicrystalline phases and the transitions
among them in the NRKHM.

The spectral and state properties of the NRKHM can be addressed by
solving the eigenvalue equation $\hat{U}|\psi_{j}\rangle=e^{-iE_{j}}|\psi_{j}\rangle$
for the Floquet operator $\hat{U}$ in Eq.~(\ref{eq:U}). Here $E_{j}$
is the quasienergy of the $j$th right eigenvector $|\psi_{j}\rangle$,
which can take complex values as $\hat{U}$ is nonunitary. For a lattice
of length $L$, there are $L$ such eigenstates indexed by $j=1,...,L$.
To check whether ${\cal PT}$-breaking transitions could happen, we
consider the maximal imaginary parts of $E$ and the density of states
(DOSs) $\rho$ with non-real quasienergies \cite{NHQCThy1,NHQCThy2,NHQCThy3},
which are defined as
\begin{alignat}{1}
	\max|{\rm Im}E| & =\max_{j\in\{1,...,L\}}(|{\rm Im}E_{j}|),\label{eq:MaxImE}\\
	\rho & =N({\rm Im}E\neq0)/L.\label{eq:DOS}
\end{alignat}
Here $N({\rm Im}E\neq0)$ means the number of states whose quasienergies
have nonvanishing imaginary parts. It is clear that once the $\max|{\rm Im}E|$
switches from zero to a finite value, eigenstates with complex quasienergies
would appear in the Floquet spectrum. A ${\cal PT}$-breaking transition
then happens in the NRKHM, which is further accompanied by the deviation
of $\rho$ from zero. Moreover, we would have $\rho\simeq1$ when
almost all eigenstates of $\hat{U}$ possess non-real quasienergies.
We can thus use $\max|{\rm Im}E|$ and $\rho$ to distinguish phases with
different spectral nature.

The localization properties of Floquet eigenstates in the NRKHM can
be characterized by their level-spacing statistics and inverse participation
ratios (IPRs) \cite{NHQCThy1,NHQCThy2,NHQCThy3}. If the set of quasienergies
${\cal E}=\{E_{j}|j=1,...,L\}$ has been sorted by their real parts,
we can identify the real level-spacing between the $j$th and the
$(j-1)$th element in the set ${\cal E}$ as $\epsilon_{j}={\rm Re}E_{j}-{\rm Re}E_{j-1}$.
The ratio between adjacent level-spacings can be defined as $g_{j}=\min(\epsilon_{j},\epsilon_{j+1})/\max(\epsilon_{j},\epsilon_{j+1})$
for $j=2,...,L-1$, where $\min(\epsilon,\epsilon')$ and $\max(\epsilon,\epsilon')$
yield the minimum and maximum of $\epsilon$ and $\epsilon'$. The
statistical feature of adjacent gap ratios (AGRs) $g_{j}$ can be
determined by the average of AGRs over all states~\cite{NHRMT1,NHRMT2,NHRMT3,NHRMT4,NHRMT5}, i.e.,
\begin{equation}
	\overline{g}=\frac{1}{L}\sum_{j}g_{j}.\label{eq:MAGR}
\end{equation}
If all bulk states are extended in the thermodynamic limit $L\rightarrow\infty$,
we would have $\overline{g}\rightarrow0$. For a phase in which all
bulk states are localized, $\overline{g}$ would instead approach
a finite constant $\overline{g}_{\max}>0$. If there is a critical
region in which extended and localized eigenstates coexist and are
separated by a mobility edge, $\overline{g}$ is nonuniversal and
take values in a range $\overline{g}\in(0,\overline{g}_{\max})$.
The behavior of $\overline{g}$ can thus be employed to distinguish
phases with different localization nature in the NRKHM. The IPR is
another direct measure of the state profiles in the system. For a
given normalized right eigenvector $|\psi_{j}\rangle=\sum_{n=1}^{L}\psi_{n}^{j}|n\rangle$
of $\hat{U}$ with quasienergy $E_{j}$ in the lattice basis, we define
its IPR as ${\rm IPR}_{j}=\sum_{n=1}^{L}|\psi_{n}^{j}|^{4}$. A conjugate
quantity, called the normalized participation ratio (NPR) can be constructed
for the state $|\psi_{j}\rangle$ as ${\rm NPR}_{j}=(\sum_{n=1}^{L}|\psi_{n}^{j}|^{4})^{-1}/L$.
If $|\psi_{j}\rangle$ happens to be a localized state (an extended
state), we would have ${\rm IPR}_{j}\rightarrow\lambda_{j}$ (${\rm IPR}_{j}\rightarrow0$)
and ${\rm NPR}_{j}\rightarrow0$ (${\rm NPR}_{j}\rightarrow1$), where
the Lyapunov exponent $\lambda_{j}$ can be a function of the quasienergy
$E_{j}$. The global localization property of the system can then
be inspected by averaging the IPRs and NPRs over all bulk states,
yielding
\begin{equation}
	{\rm IPR}_{{\rm ave}}=\frac{1}{L}\sum_{j=1}^{L}{\rm IPR}_{j},\label{eq:IPRave}
\end{equation}
\begin{equation}
	{\rm NPR}_{{\rm ave}}=\frac{1}{L}\sum_{j=1}^{L}{\rm NPR}_{j}.\label{eq:NPRave}
\end{equation}
Moreover, we introduce the minimum/maximum of IPRs and an extra quantity
$\zeta$ to capture the transitions between different phases in the
system and the presence of critical phase with mobility edge. These
quantities are explicitly defined as
\begin{equation}
	{\rm IPR}_{\max}=\max_{j\in\{1,...,L\}}({\rm IPR}_{j}),\label{eq:IPRmax}
\end{equation}
\begin{equation}
	{\rm IPR}_{\min}=\min_{j\in\{1,...,L\}}({\rm IPR}_{j}),\label{eq:IPRmin}
\end{equation}
\begin{equation}
	\zeta=\log_{10}({\rm IPR}_{{\rm ave}}\cdot{\rm NPR}_{{\rm ave}}).\label{eq:ZETA}
\end{equation}
It is clear that once the first (last) localized (extended) bulk state
appears (vanishes) in the system, the ${\rm IPR}_{\max}$ (${\rm IPR}_{\min}$)
will be found to deviate from zero in the limit $L\rightarrow\infty$.
Moreover, the value of $\zeta$ can be finite only when ${\rm IPR}_{{\rm ave}}\cdot{\rm NPR}_{{\rm ave}}\neq0$,
which means that finite amounts of extended and localized states could
survive together in the system. Therefore, the NRKHM described by
$\hat{U}$ resides in the extended, localized, or critical mobility
edge phase if ${\rm IPR}_{\max}\rightarrow0$, ${\rm IPR}_{\min}>0$,
or $\zeta$ being finite in the thermodynamic limit, respectively.

To probe the transport nature of different phases in the NRKHM, we
consider the dynamics of wavepackets \cite{NRHM2}. The stroboscopic
evolution of an initial state $|\psi(0)\rangle$ over $\ell$ driving
periods yields the state $|\tilde{\psi}(t=\ell T)\rangle=\hat{U}^{\ell}|\psi(0)\rangle$,
which is not normalized in the lattice representation for a nonunitary
$\hat{U}$. Using the normalization convention of right eigenvectors,
the normalized state after an evolution over $t=\ell T$ driving periods
takes the form $|\psi(t)\rangle=|\tilde{\psi}(t)\rangle/\sqrt{\langle\tilde{\psi}(t)|\tilde{\psi}(t)\rangle}$.
Expanding $|\psi(t)\rangle$ in the lattice basis produces the probability
amplitude $\psi_{n}(t)=\langle n|\psi(t)\rangle$ of the evolved state
at different locations of the lattice, i.e., $|\psi(t)\rangle=\sum_{n}\psi_{n}(t)|n\rangle$.
To characterize the stroboscopic dynamics of the wavepacket, we investigate
the time-dependence of its center $x(t)$, standard deviation $x_{{\rm sd}}(t)$
and averaged spreading speed $v(t)$ in the lattice, which are defined
as
\begin{equation}
	x(t)=\sum_{n}n|\psi_{n}(t)|^{2},\label{eq:X}
\end{equation}
\begin{equation}
	x_{{\rm sd}}(t)=\sqrt{\sum_{n}n^{2}|\psi_{n}(t)|^{2}-x^{2}(t)},\label{eq:Xsd}
\end{equation}
\begin{equation}
	v(t)=\frac{1}{t}\sqrt{\sum_{n}n^{2}|\psi_{n}(t)|^{2}}.\label{eq:VT}
\end{equation}
Here $t=\ell T$ refers to the stroboscopic time with $\ell\in\mathbb{Z}$.
For a localized initial state in the lattice, we expect $x(t)$ {[}$x_{{\rm sd}}(t)${]}
to stay close to its initial value in the localized phase, $x(t)\propto t$
{[}$x_{{\rm sd}}(t)\propto\sqrt{t}${]} in the extended phase due
to the nonreciprocal hopping in the system, and $x(t)$ {[}$x_{{\rm sd}}(t)${]}
to show an intervening behavior in the critical phase with mobility
edge. For the same initial state and over a long evolution time ($\ell\gg1$),
the average speed $v(t)$ should tend to vanish in the localized phase,
become finite in the intermediate phase with mobility edge due to
the presence of hopping asymmetry, and taking maximal values in the
extended phase. We can thus employ $x(t)$, $x_{{\rm sd}}(t)$ and
$v(t)$ to dynamically distinguish phases with different transport
properties in the NRKHM.

\section{Results\label{sec:Res}}

\begin{figure}
	\begin{centering}
		\includegraphics[scale=0.48]{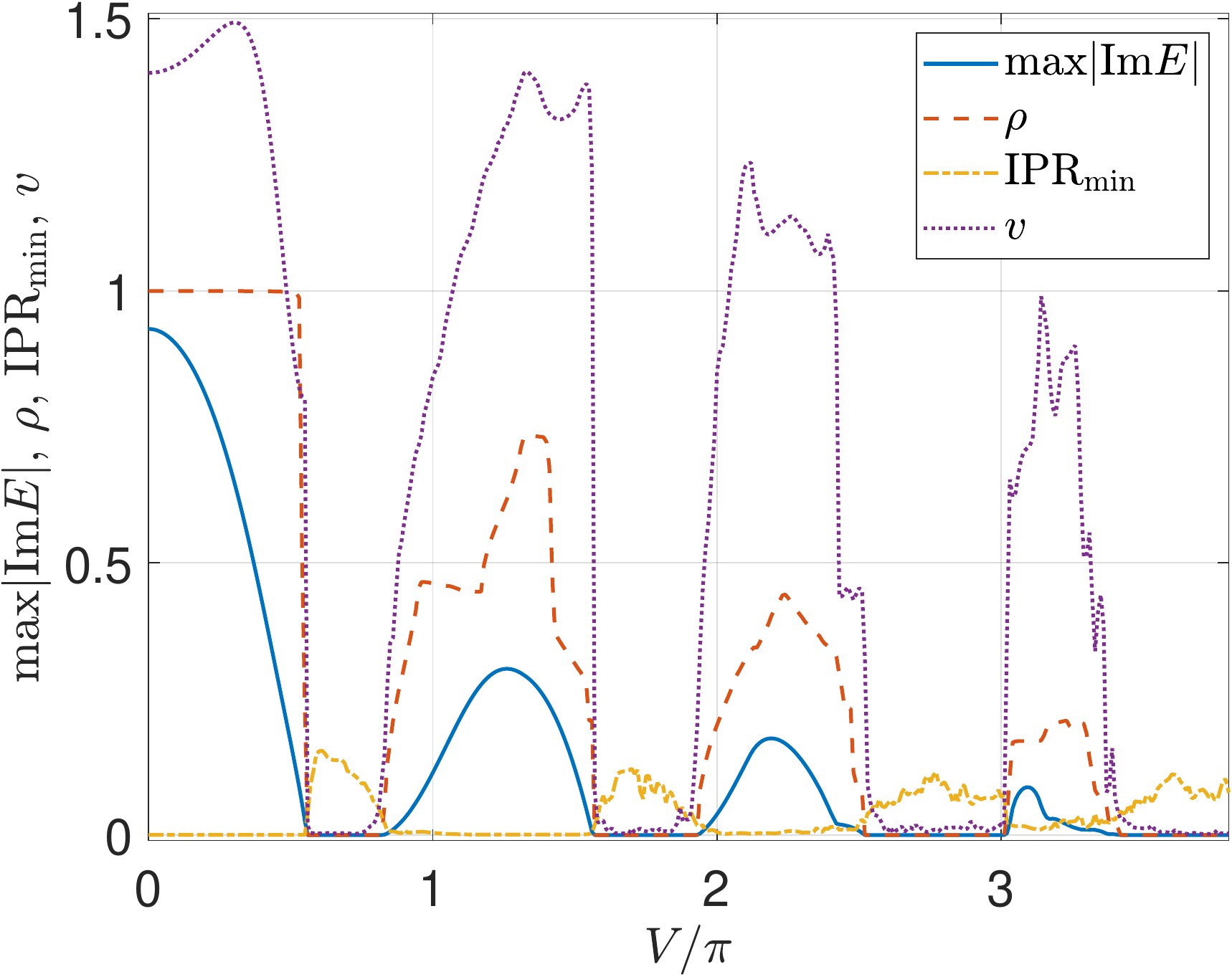}
		\par\end{centering}
	\caption{Maximal imaginary parts of quasienergy (blue solid line), DOSs with
		non-real quasienergies (red dashed line), minimum of IPRs (yellow dash-dotted
		line) and averaged spreading velocity a of wavepacket (purple dotted line)
		versus the onsite potential amplitude $V$ under the PBC. Other system
		parameters are $(J,\gamma)=(\pi/6,0.8)$. The lattice size is $L=4181$.
		The initial state in the calculation of $v=v(t)$ is chosen to be
		$|\psi(0)\rangle=\sum_{n}\delta_{n0}|n\rangle$. The average in
		Eq.~(\ref{eq:VT}) is taken over $1000$ driving periods.\label{fig:IPRmVtE}}
\end{figure}

We are now ready to uncover the spectral and localization nature of
the quasiperiodic NRKHM with the help of the tools introduced in the
last section. In Fig.~\ref{fig:IPRmVtE}, we present the maximal imaginary
parts of $E$ {[}Eq.~(\ref{eq:MaxImE}){]}, the DOSs with non-real
quasienergies {[}Eq.~(\ref{eq:DOS}){]}, the minimum of IPRs 
{[}Eq.~(\ref{eq:IPRmin}){]} and the spreading speed of an initially localized
excitation {[}Eq.~(\ref{eq:VT}){]} versus the onsite potential $V$
for a typical set of system parameters. Interestingly, we observe
complex and highly monotonous behaviors for all these quantities,
which are drastically different from the cases observed in the non-driven
\cite{NRHM1,NRHM2} or high-frequency driven \cite{NHFQC1} NRHM.
Specially, the ${\cal PT}$-symmetry is broken and the Floquet spectrum
of the system is complex at $V=0$ (i.e., the clean lattice limit).
With the increase of $V$, the quasienergy spectrum could first undergo
a ${\cal PT}$-restoring transition from complex to real. But later
it becomes complex again after a ${\cal PT}$-breaking transition
with the further increase of $V$. In the second complex-spectrum
phase, the ratio of states with non-real quasienergies $\rho$ is
finite but smaller than $1$, which means that states with real and
complex quasienergies coexist in this phase. With the further increase
of $V$, the NRKHM encounters a sequence of ${\cal PT}$-restoring
and ${\cal PT}$-breaking transitions, while its quasienergy spectrum
alternates between purely real and partially complex in different
parameter regions. More interestingly, the ${\rm IPR}_{\min}$ is
pinned to zero in the phases with complex quasienergies and deviates from zero
whenever the spectra become real. This suggests that all eigenstates
in real-spectrum phases of the NRKHM are localized, whereas extended
eigenstates emerge whenever non-real quasienergies appear in the Floquet
spectrum. The connection between the realness of Floquet spectrum
and the localization nature of states is further confirmed by investigating
the spreading of a localized initial wavepacket, whose average speed
is finite in the regions with $\max|{\rm Im}E|>0$, ${\rm IPR}_{\min}\simeq0$,
but vanishes in the domains with $\max|{\rm Im}E|=0$, ${\rm IPR}_{\min}>0$,
yielding the phenomenon of dynamical localization in real-spectrum
phases of the NRKHM.

\begin{figure*}[ht]
	\begin{centering}
		\includegraphics[scale=0.5]{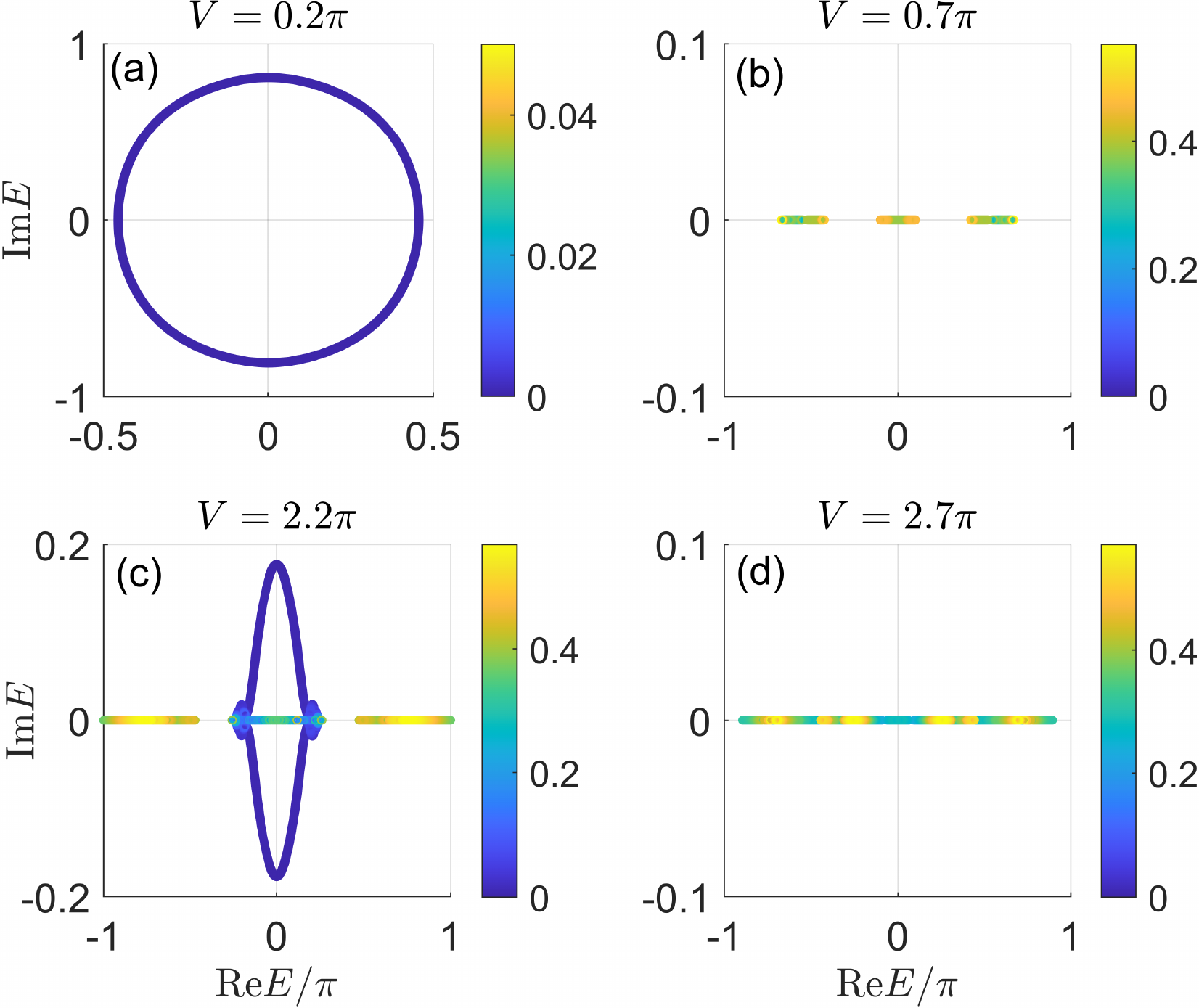}\includegraphics[scale=0.5]{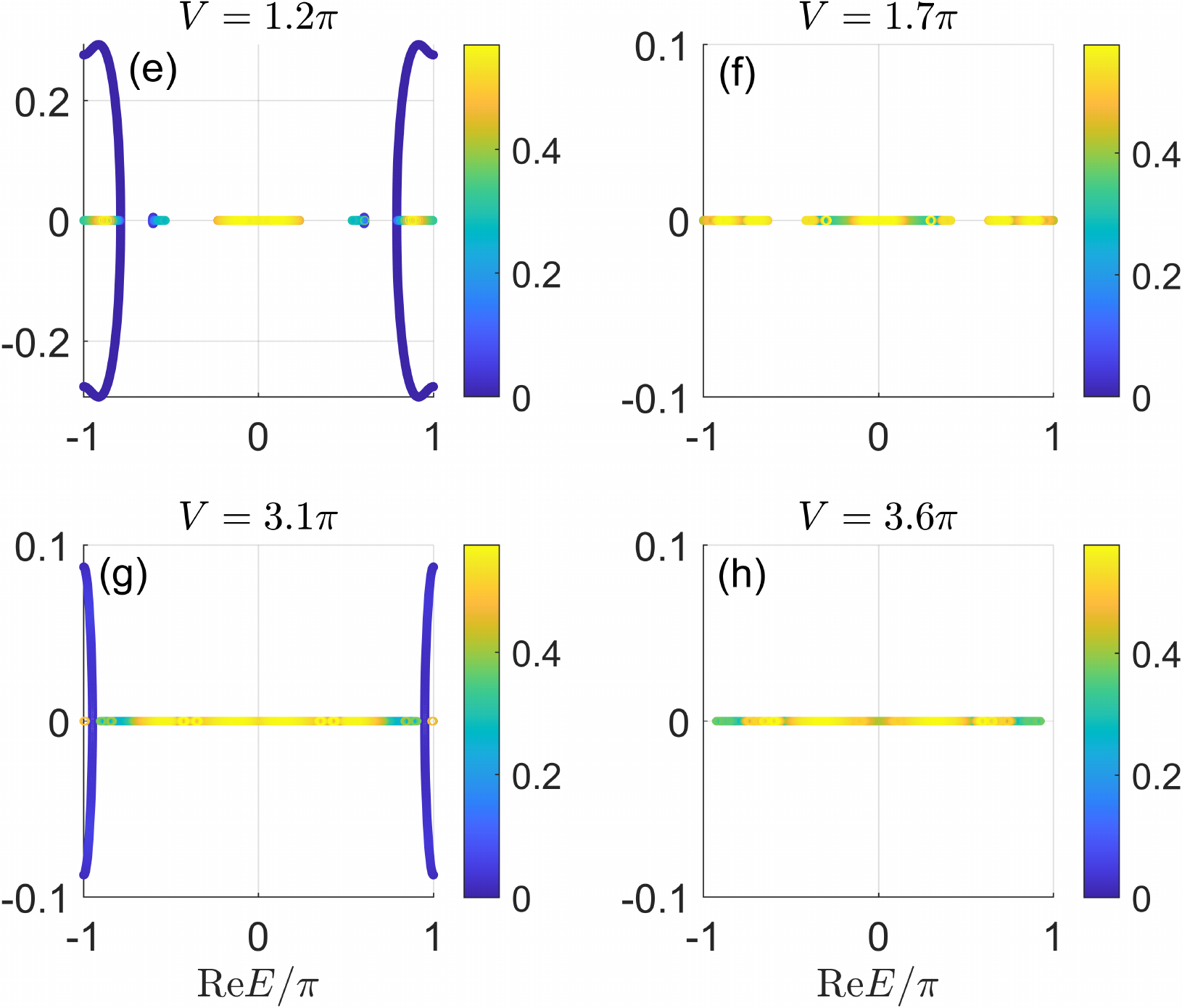}
		\par\end{centering}
	\caption{Floquet spectrum $E={\rm Re}E+i{\rm Im}E$ on the complex plane (in circles) and IPRs (in
		color scale) of the NRKHM under PBC for eight typical cases. The amplitude
		$V$ of onsite quasiperiodic potential is shown in the caption of
		each figure panel. Other system parameters are set as $(J,\gamma)=(\pi/6,0.8)$.
		The length of lattice is chosen to be $L=4181$ for all panels. \label{fig:E}}
\end{figure*}

To further clarify the relationship between the structure of spectrum
and the states' localization property in the NRKHM, we pick out a set
of typical parameters from Fig.~\ref{fig:IPRmVtE}, and show the corresponding
quasienergy spectra on the complex plane together with the IPRs of
all Floquet eigenstates in Fig.~\ref{fig:E}. With a weak quasiperiodic
potential ($V=0.2\pi$), all states are found to be extended (${\rm IPR}\simeq0$)
and taking non-real quasienergies with the hopping asymmetry $\gamma\neq0$,
as shown in Fig.~\ref{fig:E}(a). For parameters chosen within the
real-spectrum phases of Fig.~\ref{fig:IPRmVtE} ($V=0.7\pi,1.7\pi,2.7\pi,3.6\pi$),
we instead observe localization for all Floquet eigenstates (${\rm IPR}>0$)
as shown in Figs.~\ref{fig:E}(b), \ref{fig:E}(d), \ref{fig:E}(f),
and \ref{fig:E}(h). For parameters taken in the second ($V=1.2\pi$),
third ($V=2.2\pi$), and fourth ($V=3.1\pi$) mixed-spectrum phases
of Fig.~\ref{fig:IPRmVtE}, we find the coexistence of real and non-real
quasienergies in different regions of the spectrum. Moreover, eigenstates
with real (non-real) quasienergies are found to be localized (extended)
in all the three cases, as shown in Figs.~\ref{fig:E}(c), \ref{fig:E}(e),
and \ref{fig:E}(g). The mixed-spectrum phases appear at larger $V$
are thus coincide with the critical phases of NRKHM, in which localized
real-quasienergy eigenstates and extended non-real-quasienergy eigenstates are
separated by mobility edges on the complex plane. Therefore, as a
result of the competition among driving, non-Hermiticity and disorder,
the quasiperiodic NRKHM possesses at least three distinct phases,
i.e., an extended phase with purely complex spectrum, a localized
phase with real spectrum, and a critical phase wherein states with
real and non-real quasienergies coexist and are separated by mobility
edges.

\begin{figure*}[ht]
	\begin{centering}
		\includegraphics[scale=0.5]{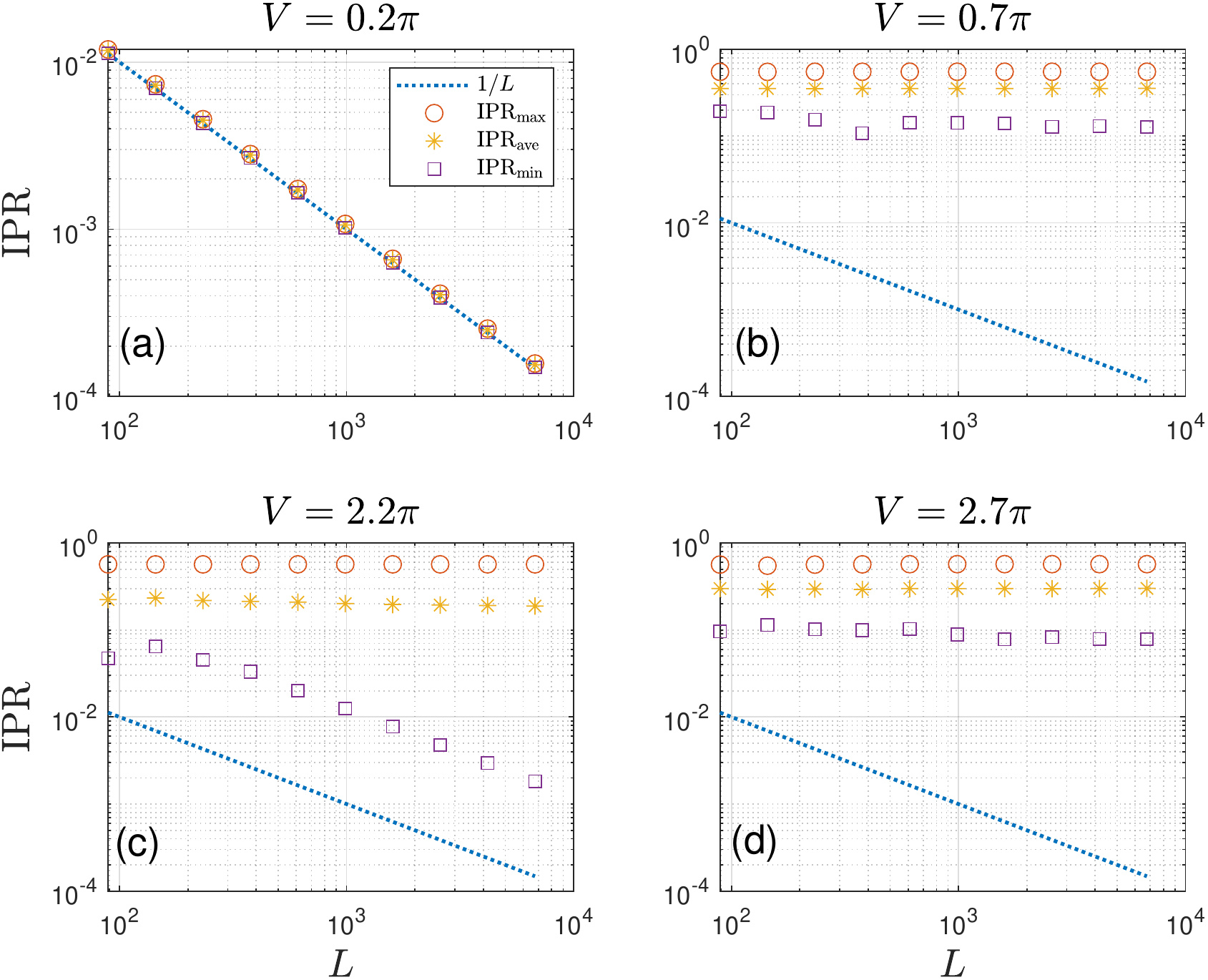}\includegraphics[scale=0.5]{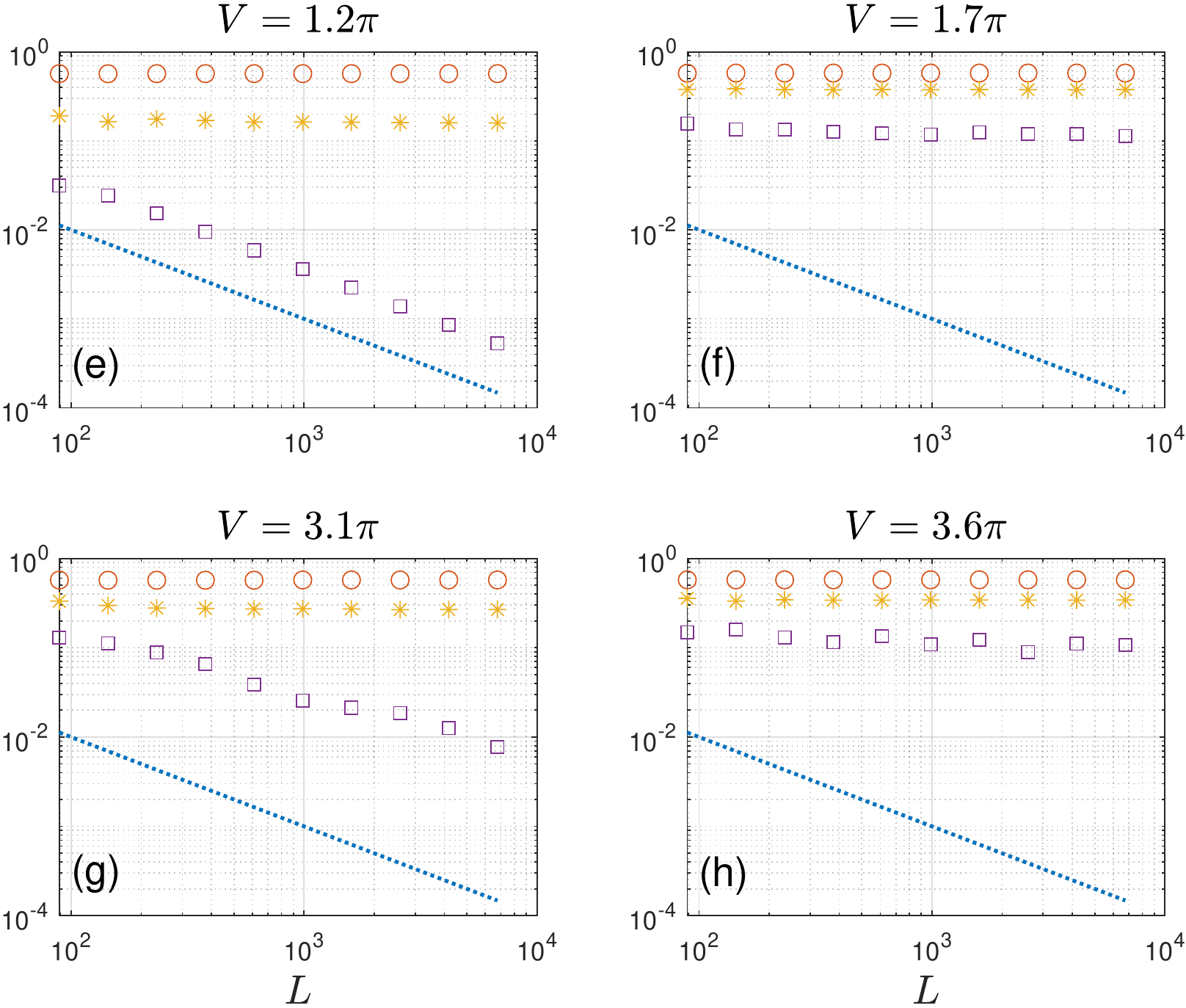}
		\par\end{centering}
	\caption{Scaling of the IPRs of right eigenvectors versus the system size $L$ under the PBC, shown in loglog plot.
		In all panels, the dotted lines, circles, stars and squares denote
		the values of $1/L$, ${\rm IPR}_{\max}$, ${\rm IPR}_{{\rm ave}}$ and
		${\rm IPR}_{\min}$ at different lattice sizes $L$ for $L=89,144,233,377,610,987,1597,2584,4181,6765$.
		The amplitude $V$ of the onsite potential is given in the caption of
		each figure panel. Other system parameters are set as $(J,\gamma)=(\pi/6,0.8)$.
		\label{fig:IPRvsL}}
\end{figure*}

To consolidate the presence of the three possible phases in the thermodynamic
limit, we show in Fig.~\ref{fig:IPRvsL} the scaling of IPRs versus
the lattice size $L$ for the eight representative cases of Fig.~\ref{fig:E}
under the PBC. As expected, we find that the ${\rm IPR}_{\max}$ 
{[}Eq.~(\ref{eq:IPRmax}){]}, ${\rm IPR}_{{\rm ave}}$ {[}Eq.~(\ref{eq:IPRave}){]}
and ${\rm IPR}_{\min}$ {[}Eq.~(\ref{eq:IPRmin}){]} are all $\propto L^{-1}$
with the increase of $L$ in Fig.~\ref{fig:IPRvsL}(a), implying that
all eigenstates are extended in the phase with purely complex spectrum.
Comparatively, for parameters taken in the real-spectrum phases, we
find almost no changes in ${\rm IPR}_{\max}$, ${\rm IPR}_{{\rm ave}}$
and ${\rm IPR}_{\min}$ with the increase of $L$ as shown in 
Figs.~\ref{fig:IPRvsL}(b), \ref{fig:IPRvsL}(d), \ref{fig:IPRvsL}(f),
and \ref{fig:IPRvsL}(h), meaning that all states in these cases are
localized. For parameters taken in the mixed-spectrum phases, the
${\rm IPR}_{\max}$ is found to be independent of $L$, whereas the
${\rm IPR}_{\min}$ is inversely proportional to $L$ as shown in
Figs.~\ref{fig:IPRvsL}(c), \ref{fig:IPRvsL}(e), and \ref{fig:IPRvsL}(g),
verifying that localized and extended states coexist in the system
in these cases. The real-spectrum localized, complex-spectrum extended
and mixed-spectrum critical mobility edge phases of the NRKHM are
thus expected to survive in the thermodynamic limit $L\rightarrow\infty$.

\begin{figure}
	\begin{centering}
		\includegraphics[scale=0.48]{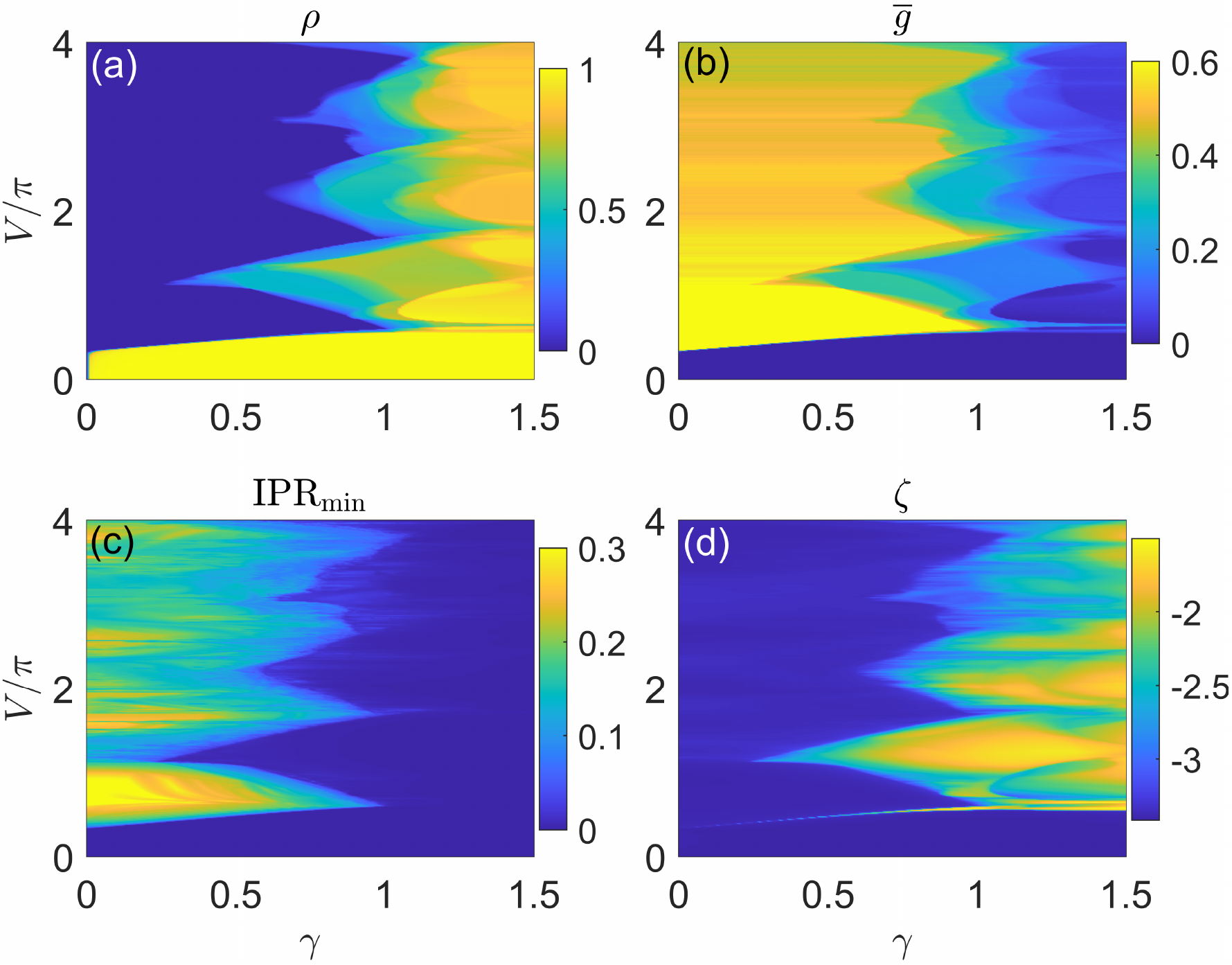}
		\par\end{centering}
	\caption{The DOSs $\rho$, averaged AGRs $\overline{g}$, minimum of IPRs ${\rm IPR}_{\min}$
		and probe of critical mobility edge phase $\zeta$ versus the hopping
		asymmetry parameter $\gamma$ and onsite potential amplitude $V$ under the PBC.
		The uniform part of hopping amplitude is $J=\pi/6$ and the length
		of lattice is $L=2584$ for all calculations. \label{fig:PhsDiag}}
\end{figure}

To acquire a more complete understanding of the phases and transitions
in the NRKHM, we present the DOSs {[}Eq.~(\ref{eq:DOS}){]}, the averaged
AGRs {[}Eq.~(\ref{eq:MAGR}){]}, the minimum of IPRs {[}Eq.~(\ref{eq:IPRmin}){]}
and the measure of critical mobility edge phase $\zeta$ {[}Eq.~(\ref{eq:ZETA}){]}
versus the amount of hopping asymmetry $\gamma$ and the onsite potential
$V$ in Fig.~\ref{fig:PhsDiag}. The consistency of the results among
the four panels of Fig.~\ref{fig:PhsDiag} clearly suggests the presence
of an extended phase (with $\rho\simeq1$, $\overline{g}\simeq0$,
${\rm IPR}_{\min}\simeq0$, and $\zeta\rightarrow-\infty$), a localized
phase (with $\rho\simeq0$, $\overline{g}\simeq0.6$, ${\rm IPR}_{\min}>0$,
and $\zeta\rightarrow-\infty$) and a critical mobility edge phase
(with $0<\rho<1$, $0\lesssim\overline{g}\lesssim0.6$, ${\rm IPR}_{\min}>0$,
and $\zeta$ finite) in the NRKHM. When the non-Hermitian parameter
$\gamma$ is small, the change of the quasiperiodic potential amplitude
$V$ can only cause a transition of the system from a complex-spectrum
extended phase to a real-spectrum localized phase, which is similar
to what happens in the non-driven NRHM \cite{NRHM1,NRHM2}. When the
hopping asymmetry $\gamma$ is large enough, non-Hermitian effects
dominant. In this region, we only find the transition from a complex-spectrum
extended phase to a mixed-spectrum critical phase in Fig.~\ref{fig:PhsDiag}.
This observation suggests that real-quasienergy extended states could
persist in the NRKHM over a broad range of quasiperiodic potential
amplitude $V$ at strong non-Hermiticity. In the intermediate range
of $\gamma$, however, we find reentrant spectral and localization
transitions between real-spectrum localized and mixed-spectrum critical
phases with the increase of $V$. This observation demonstrates unambiguously
that the rich phase and transition patterns in the NRKHM are indeed
originated from the interplay among three nontrivial effects when
they are comparable, i.e., the non-Hermiticity, Floquet driving and
spatial quasiperiodicity. The original phase diagram of the NRHM \cite{NRHM1,NRHM2}
gets most strongly modified when both the hopping nonreciprocity and
the driving field reach sufficient strengths.

\begin{figure*}[ht]
	\begin{centering}
		\includegraphics[scale=0.5]{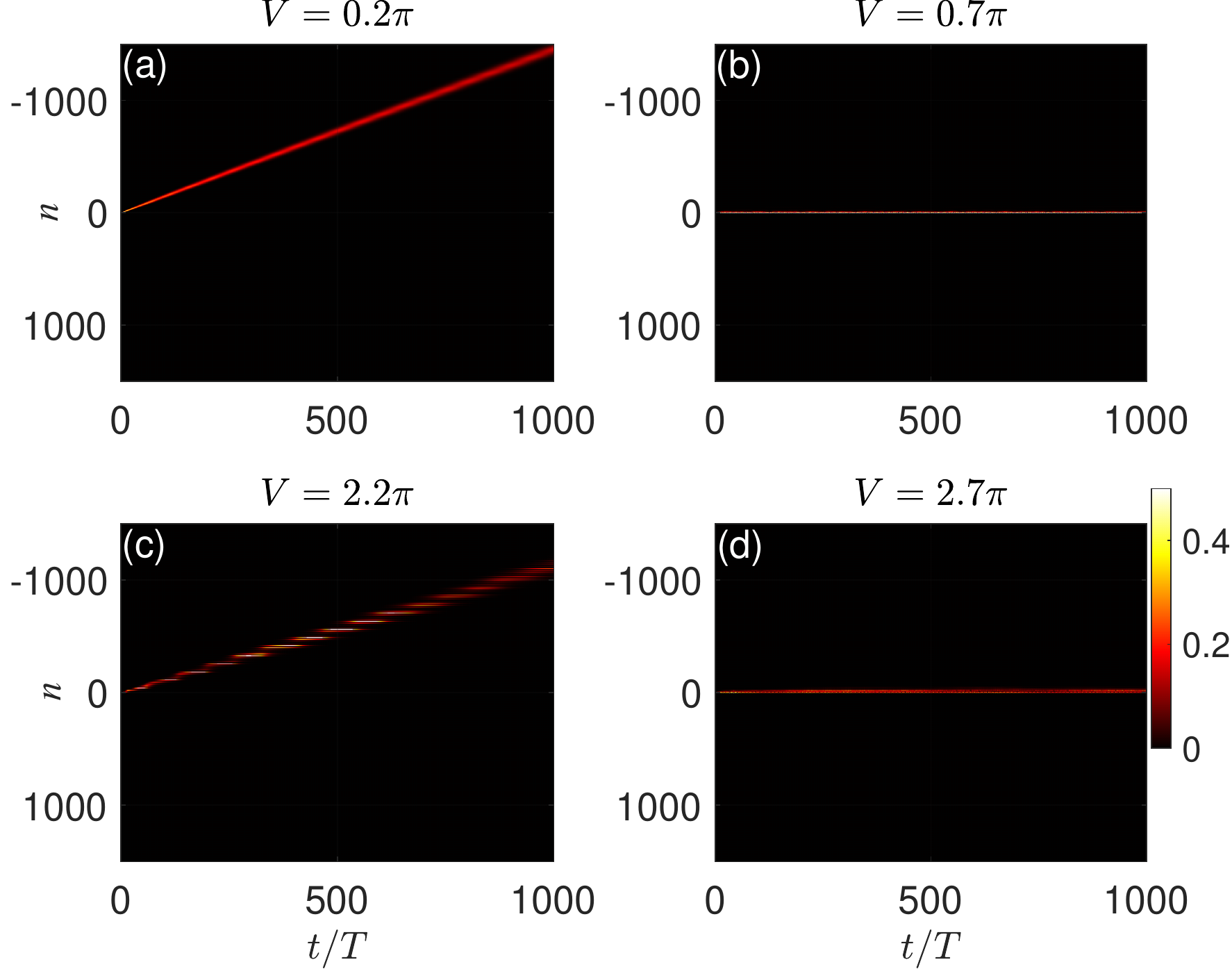}\includegraphics[scale=0.5]{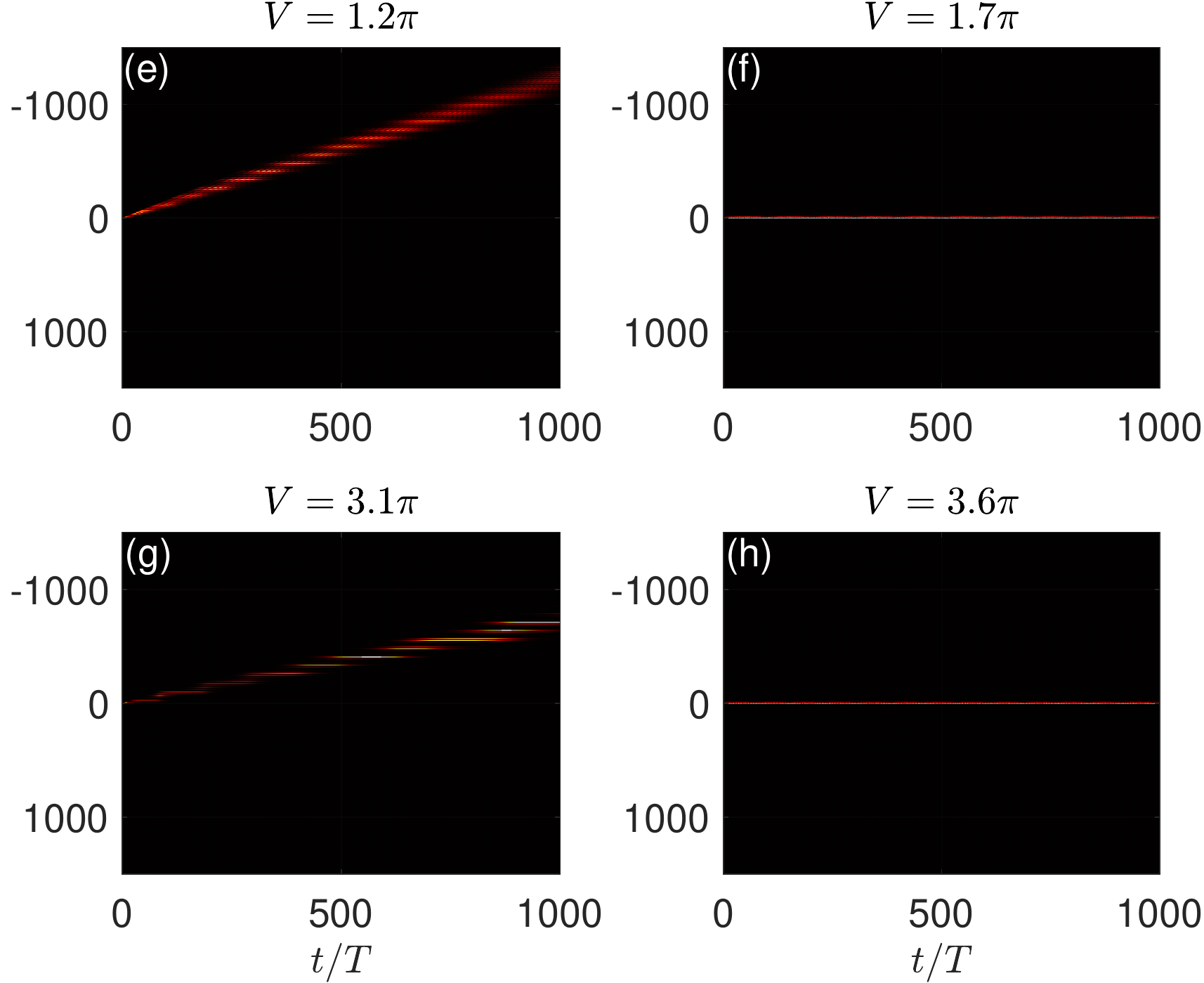}
		\par\end{centering}
	\caption{Propagation of an initially localized wavepacket $|\psi(0)\rangle=\sum_{n}\delta_{n0}|n\rangle$
		in the NRKHM under the PBC. The color scale denotes the absolute amplitude
		of the evolving state in the lattice at the time $t=\ell T$ for $\ell=1,2,...,999,1000$.
		All panels share the same color bar.
		The values of onsite potential amplitude $V$ are given in corresponding
		figure captions. Other system parameters are $(J,\gamma)=(\pi/6,0.8)$.
		The length of lattice is $L=4181$ and the lattice site index (vertical axis)
		take values from $n=-2090$ to $n=2090$. \label{fig:Pt}}
\end{figure*}

The presence of extended, localized and critical phases in the NRKHM
has immediate dynamical implications, which could help us to detect
and differentiate them in experiments. To achieve this goal, we consider
the dynamics of a single-site excitation in the lattice under the PBC,
which is described by the theory introduced in Sec.~\ref{sec:Met}.
The spatial distribution of probability amplitude $|\psi_n(t)|=|\langle n|\psi(t)\rangle|$ of wavepackets at different stroboscopic
times $t=\ell T$ ($\ell\in\mathbb{Z}$) in the system are shown in
Fig.~\ref{fig:Pt} for the eight exemplary cases reported in previous
spectrum and localization studies. In Fig.~\ref{fig:Pt}(a), the system
is in an extended phase and the wavepacket shows a unidirectional
transport with a limited width of spreading. The former is due to the asymmetric
hopping of the lattice, i.e., tunneling from the right to left lattice sites is stronger
then the opposite for a positive $\gamma$. In Figs.~\ref{fig:Pt}(b),
\ref{fig:Pt}(d), \ref{fig:Pt}(f) and \ref{fig:Pt}(h), the system
is set in real-spectrum localized phases, and the excitation tends
to be localized around its initial position up to small oscillations.
All states in the system in these cases are thus not only spatially
but also dynamically localized. For systems prepared in the critical
mobility edge phase, the initial excitation is still found to be able to propagate
unidirectionally with a well-localized profile during the evolution,
as shown in Figs.~\ref{fig:Pt}(c), \ref{fig:Pt}(e), and \ref{fig:Pt}(g).
However, two differences are observed compared with the case shown
in Fig.~\ref{fig:Pt}(a). First, the distance between the final and
initial locations of the wavepacket in the critical phase is smaller
then that in the extended phase. This means that the excitation has
a smaller propagation speed on average when evolving in the critical
phase, which is due to the presence of localized Floquet eigenmodes
that could hinder its transport there. Second, the location of wavepacket
shows a smooth and linear growth with time in the extended phase.
Whereas in the critical phase, the location of excitation could be
pinned in space for some time during the evolution, and then grows
through tunneling to farther sites within a relatively short time
window, which is a unique phenomenon in non-Hermitian transport.

\begin{figure*}[ht]
	\begin{centering}
		\includegraphics[scale=0.5]{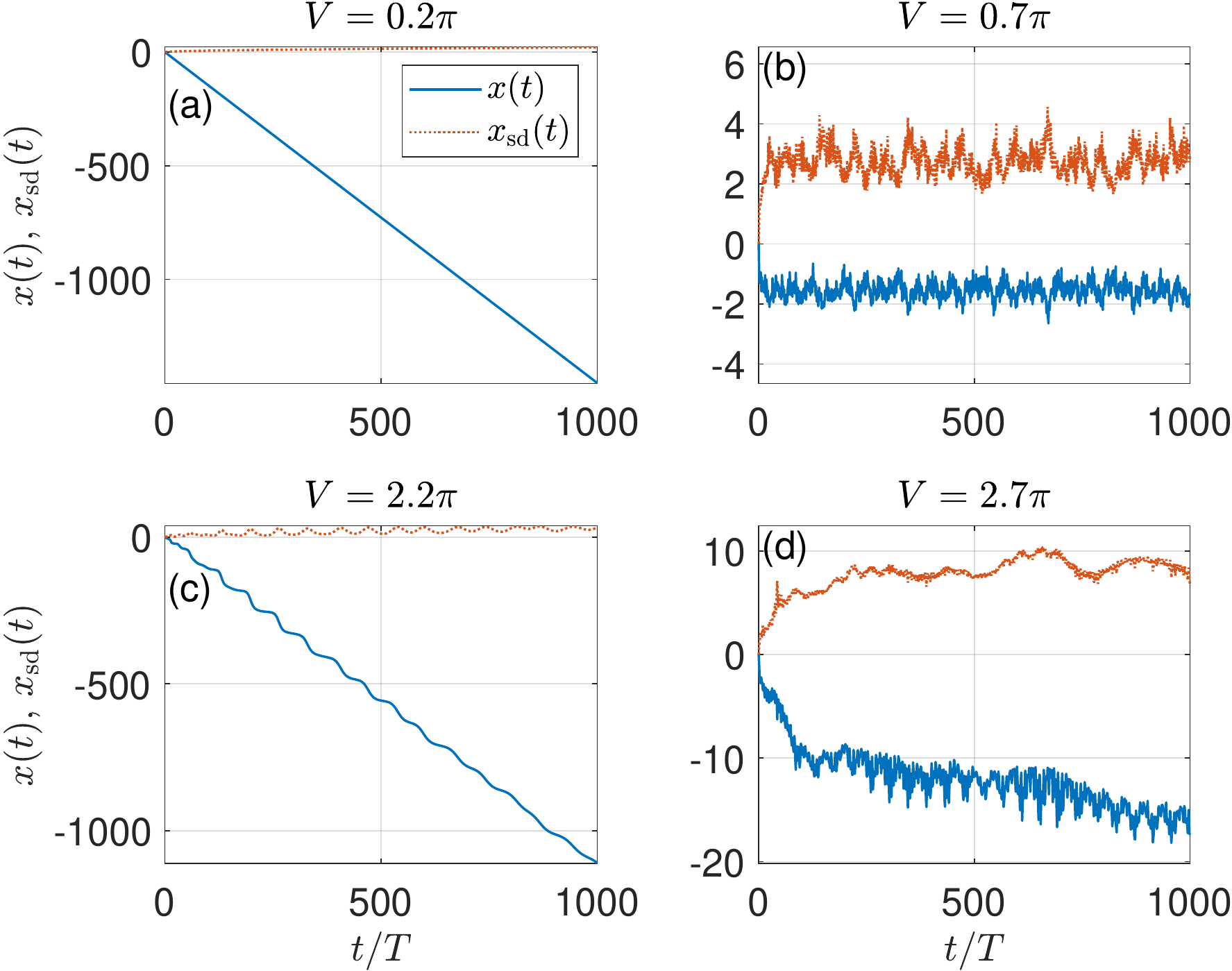}\includegraphics[scale=0.5]{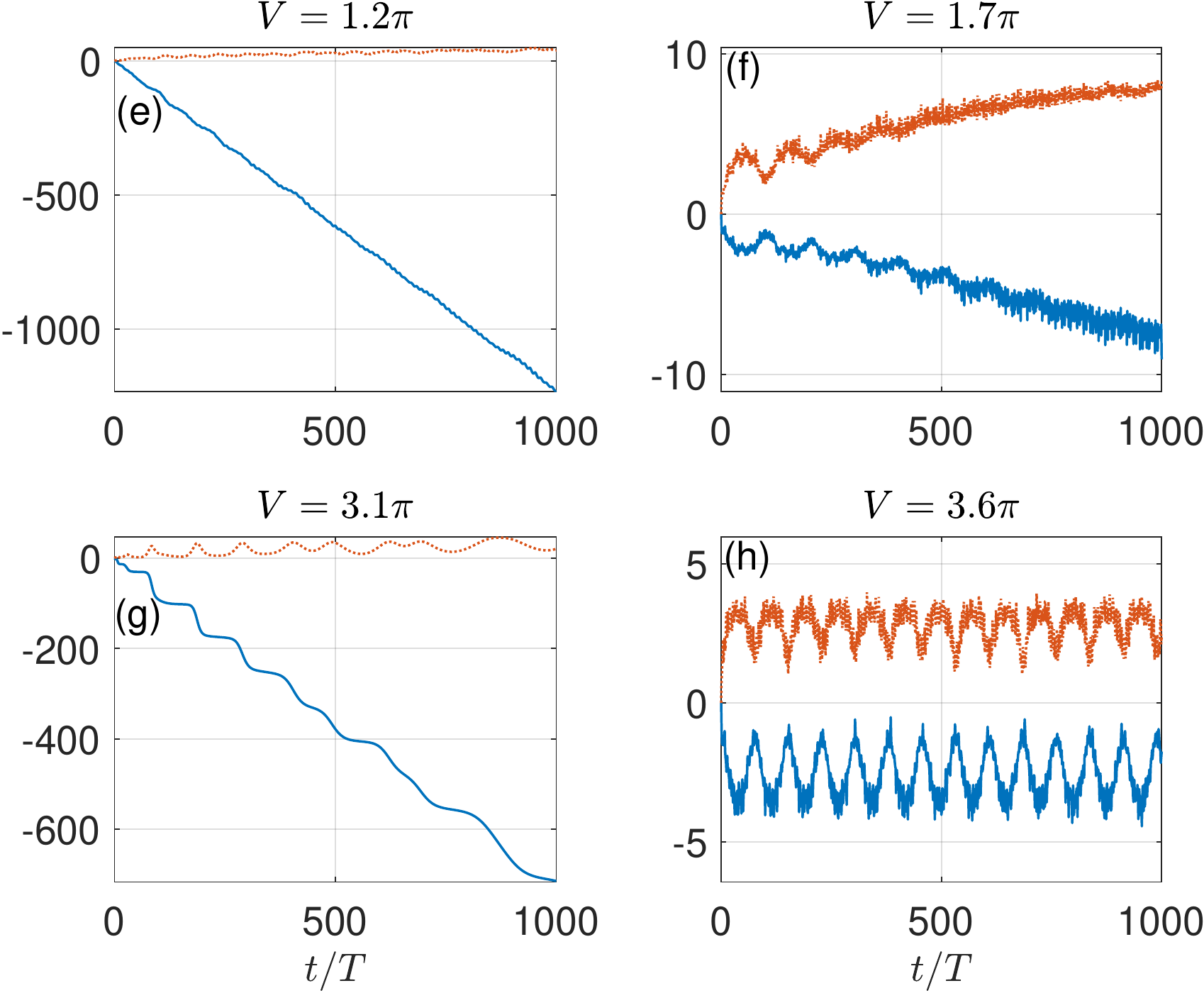}
		\par\end{centering}
	\caption{Time-dependence of the mean position $x(t)$ and standard deviation
		$x_{{\rm sd}}(t)$ of a localized initial excitation $|\psi(0)\rangle=\sum_{n}\delta_{n0}|n\rangle$
		at the center of the NRKHM under the PBC. The lattice size and system parameters
		are the same as those used in Fig.~\ref{fig:Pt}. \label{fig:Xt}}
\end{figure*}

For completeness, we present the time-dependence of the center 
{[}Eq.~(\ref{eq:X}){]} and standard deviation {[}Eq.~(\ref{eq:Xsd}){]}
of the same initial excitation for the eight parameter sets in 
Fig.~\ref{fig:Xt}. The results confirm our observations of the wavepacket
dynamics for the corresponding cases in Fig.~\ref{fig:Pt}. Put together,
the markedly different dynamical behaviors of wavepackets in distinct
parameter regions of the NRKHM could indeed provide us with a means
to probe and distinguish the extended, localized and critical phases
therein.

\section{Discussion}\label{Dis}

We now discuss more about the physical origin of the rich phase patterns and multiple reentrant transitions in the NRKHM. 
In general, ${\cal PT}$ transitions in non-Hermitian quasicrystals could appear due to the competition between Hermitian and non-Hermitian effects. Without Floquet driving fields, the impacts of Hermitian terms (the onsite potential in our case) or the non-Hermitian terms (the asymmetric hopping in our case) usually change monotonically with the increase of their control parameters ($V$ for the Hermitian and $\gamma$ for the non-Hermitian term in our model). In this case, there is only a single ${\cal PT}$ transition when the strengths of the control parameters of Hermitian (or non-Hermitian) terms dominate (exceptions exist in systems with dimerized lattice structures \cite{NHQCThy3}). With Floquet driving fields, the impacts of Hermitian and non-Hermitian terms may not change monotonically with the increase of their control parameters. Their competitions in the presence of this non-monotonicity could then result in the multiple ${\cal PT}$ and reentrant localization transitions in non-Hermitian quasicrystals.

In our system, the Floquet operator in Eq.~(\ref{eq:U}) can be equivalently written as $\hat{U}=e^{-i\hat{H}_{2}T/\hbar}e^{-i\hat{H}_{1}T/\hbar}=e^{-i\hat{H}_{{\rm eff}}T/\hbar}$,
where $\hat{H}_{1}=\hat{K}T_{1}/T$ and $\hat{H}_{2}=\hat{V}T_{2}/T$
represent rescaled hopping and onsite potential terms in the quenched
Hamiltonian. According to the Baker-Campbell-Hausdorff (BCH) formula \cite{TongPRB2013},
the effective Hamiltonian $\hat{H}_{{\rm eff}}$ can be expanded into
a series as
\begin{equation}
	\hat{H}_{{\rm eff}}=\hat{H}_{1}+\hat{H}_{2}+\frac{iT}{2}[\hat{H}_{1},\hat{H}_{2}]+\frac{T^{2}}{12\hbar^{2}}[\hat{H}_{2}-\hat{H}_{1},[\hat{H}_{1},\hat{H}_{2}]]+...\label{eq:Heff}
\end{equation}
The combination of the first two terms on the right-hand-side (RHS)
of Eq.~(\ref{eq:Heff}) just yields the static NRHM, which does not
hold any signatures of critical mobility edge phases \cite{NRHM1,NRHM2}.
However, the third, fourth and higher order terms in the expansion
on the RHS of Eq.~(\ref{eq:Heff}) contain commutators between the
onsite potential and nearest-neighbor hopping terms, which can yield
asymmetric hoppings beyond nearest-neighbor sites in the driven lattice.
With quasiperiodic disorder, these long-range hopping terms made possible
by the driving field are the usual origin of mobility edges in both
tight-binding lattices and continuous models \cite{ME1,ME2,ME3}.
In Appendix \ref{app1}, we give a brief discussion of the system at the level of $\hat{H}_{{\rm eff}}$ in Eq.~(\ref{eq:Heff}) by truncating the series at the order $T^2$. The results indeed provide evidence for the presence of reentrant localization transitions in the NRKHM.
Second, on each lattice site, the magnitude of $\hat{V}$ in $\hat{H}_{2}$
is defined modulus $2\pi$. Therefore, the contribution of each onsite
term $\mathsf{V}_{n}=VT_{2}\cos(2\pi\alpha n)/\hbar$ to the diagonal
elements of $e^{-i\hat{H}_{2}T/\hbar}$ and thus to $\hat{U}$ is
$2\pi$-periodic. This fact, in combination with other quasiperiodic
terms generated by the commutators in Eq.~(\ref{eq:Heff}) allow the
effective strength of correlated disorder to depend on $V$ in a non-monotonic
and oscillatory manner. Such a ``quasiperiodic'' $V$-dependence,
which is also enabled by the driving field, finally causes the reentrant
transitions between real-spectrum localized phases and mixed-spectrum
critical phases in the NRKHM. Notably, the range of onsite potential
$V$ considered in our study clearly include cases of strong and resonant
drivings. Our findings thus go beyond previous results focusing on
high-frequency modulation schemes \cite{NHFQC1,NHFQC2}.

Specially, in Ref.~\cite{NHFQC1}, a harmonic driving force of the form $\sum_{n}nK\cos(\omega t)$ was applied to non-Hermitian quasicrystals and the high-frequency limit was taken. The hopping amplitude $J$ is dressed by the Floquet field into the form $J_{\rm eff}=J{\cal J}_{0}(K/\omega)$, with $K$ ($\omega$) the driving amplitude (frequency) and ${\cal J}_{0}$ the Bessel function. Due to the non-monotonic dependence of ${\cal J}_{0}(K/\omega)$ on $K/\omega$, $J_{\rm eff}$ can also change with the driving parameter $K/\omega$ in a non-monotonous way, and multiple ${\cal PT}$ transitions could appear. Compared with the present work, a main difference is that in Ref.~\cite{NHFQC1}, the possible number of non-Hermitian quasicrystal phases is not modified by the driving. The driving field mainly achieves the control of the transition boundary between different phases. However, in the present work, the strong and near-resonant driving field creates multiple critical phases with mobility edges, which are absent when the driving field is switched off or the high-frequency limit is taken. The driving field further induces transitions between the newly formed critical phases and other phases with different localization nature, which are also absent in the non-driven setup. 
Therefore, the driving field in the present work plays a more non-perturbative and nontrivial role compared with what it did in Ref.~\cite{NHFQC1}.
Besides, the driving is introduced here as time-periodic quenches, whose effect may be viewed as the superposition of many harmonic drives with different frequencies and amplitudes \cite{YapPRB2018}. The hybrid nature of the drive and the long-range coupling it induced allow us to have richer patterns of non-Hermitian Floquet quasicrystal phases in the NRKHM.

\section{Summary\label{sec:Sum}}

In this work, we found that the interplay among time-periodic driving,
hopping nonreciprocity and correlated disorder could induce critical phases with mobility edges, multiple ${\cal PT}$ transitions and reentrant localization
transitions in 1D lattices. The corresponding non-Hermitian Floquet
quasicrystals possess rich phase patterns with distinct spectral and transport
nature, which are characterized by the level-spacing statistics, inverse
participation ratios and wavepacket dynamics. The reentrant transitions
among localized and critical phases in the system are physically originated
from the $2\pi$-periodicity in the amplitude of the periodically
modulated onsite potential and the effective long-range hopping generated
by high-order commutators between kinetic and potential energy terms.
Our results thus establish in aperiodic systems a unique class of
non-Hermitian Floquet matter, which holds rich and highly tunable
spectral, dynamical and localization properties. 

Two implications of our results deserve to be emphasized. First, the multiple ${\cal PT}$ transitions and reentrant localization transitions in our system emerge only when $\gamma\neq0$ and the quasiperiodic quench strength $V$ is relatively large. These transitions are thus induced by the strong interplay between Floquet driving fields and non-Hermitian effects in a disordered system. To the best of our knowledge, the multiple and reentrant transitions originated from the corporation between these two nonequilibrium knobs were not revealed in previous studies of non-Hermitian quasicrystals or Floquet states of matter, or largely overlooked in related work \cite{NHFQC2,NHFPhot5,SatijaPRE2002}. Second, the multiple lobes appearing at finite hopping nonreciprocity and quasiperiodic quench potential contain both extended and localized states. These lobes represent critical phases with quasienergy-dependent mobility edges (see Fig.~\ref{fig:IPRvsL}) in non-Hermitian Floquet quasicrystals, instead of describing localized regimes. Notably, these critical phases are absent when either the driving field or the non-Hermitian effect is switched off. Therefore, the emergence of these critical phases is a unique outcome of the nontrivial collaboration among driving, disorder and non-Hermitian effects in our system. This discovery goes beyond the previous finding \cite{NHFQC1}, in which high-frequency drivings only modulate the hopping amplitude and lead to the deformation of phase diagram without changing the number of possible phases that could appear. Our discovery thus uncovered that exceeding the high-frequency regime, the driving field could create new non-Hermitian quasicrystal phases beyond the underlying nondriven system.

In future work, it would be interesting to consider Floquet quasicrystals with lattice dimerization \cite{DimerQC1,DimerQC2,DimerQC3}
(one such example is treated briefly in Appendix \ref{app2}),
in higher spatial
dimensions and in systems with other non-Hermitian effects like onsite
gain and loss. The impact of interactions and the possible appearance
of non-Hermitian Floquet many-body localized phases in quasiperiodic
systems also deserve more thorough explorations.

\begin{acknowledgments}
L.Z. is supported by the National Natural Science Foundation of China (Grant No.~11905211), the Young Talents Project at Ocean University of China (Grant No.~861801013196), and the Applied Research Project of Postdoctoral Fellows in Qingdao (Grant No.~861905040009).
\end{acknowledgments}

\appendix

\section{Phase transitions described by ${\hat H}_{\rm eff}$}\label{app1}

\begin{figure}
	\begin{centering}
		\includegraphics[scale=0.48]{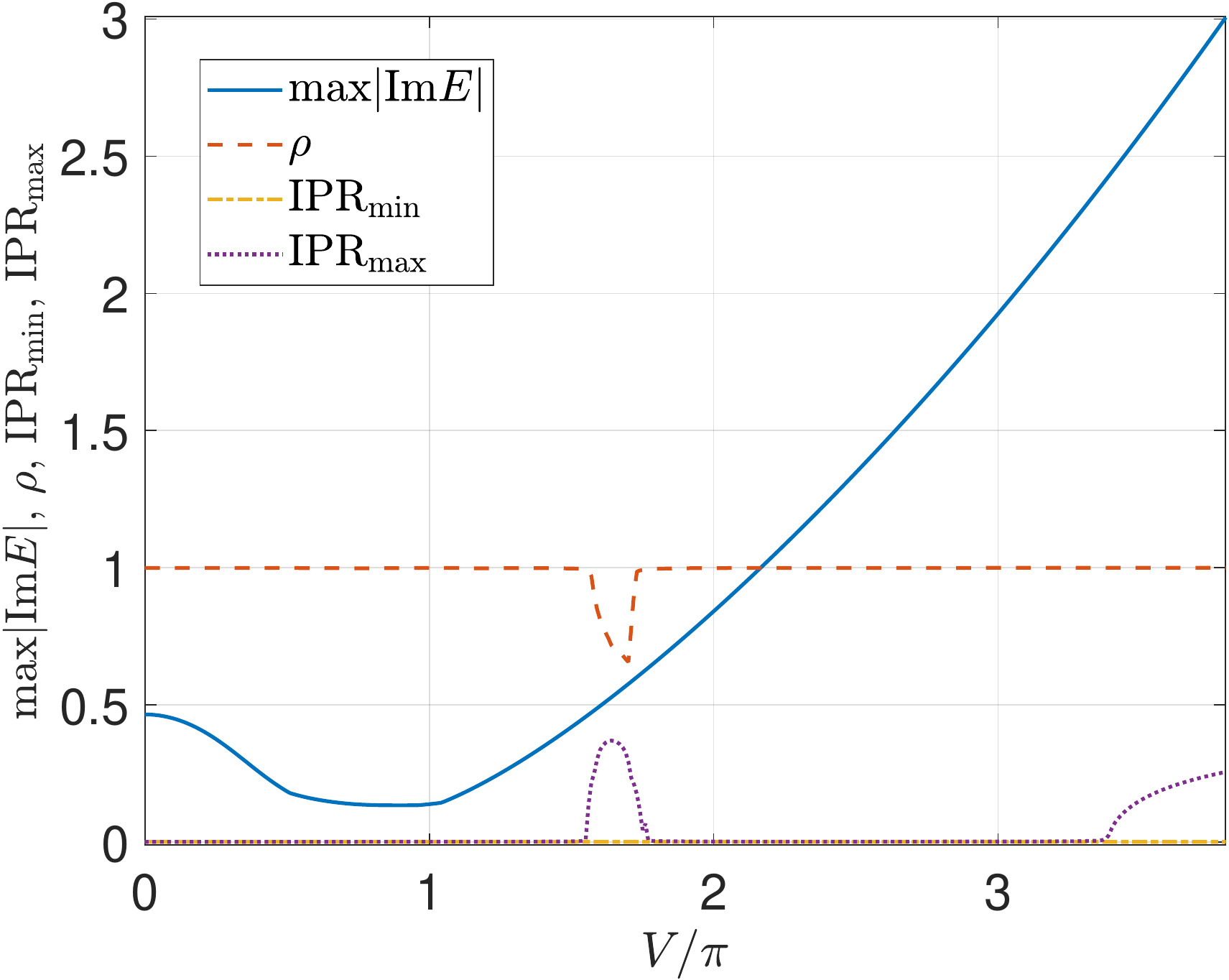}
		\par\end{centering}
	\caption{Maximal imaginary parts of quasienergy (blue solid line), DOSs with non-real quasienergies (red dashed line), minimum and maximum of IPRs (yellow dash-dotted and purple dotted lines) of the ${\hat H}_{\rm eff}$ versus the onsite potential amplitude $V$ under the PBC. Other system parameters are $(J,\gamma)=(\pi/6,0.8)$. The lattice size is $L=4181$.}\label{fig:IPRmVtE-Heff}
\end{figure}

\begin{figure}
	\begin{centering}
		\includegraphics[scale=0.48]{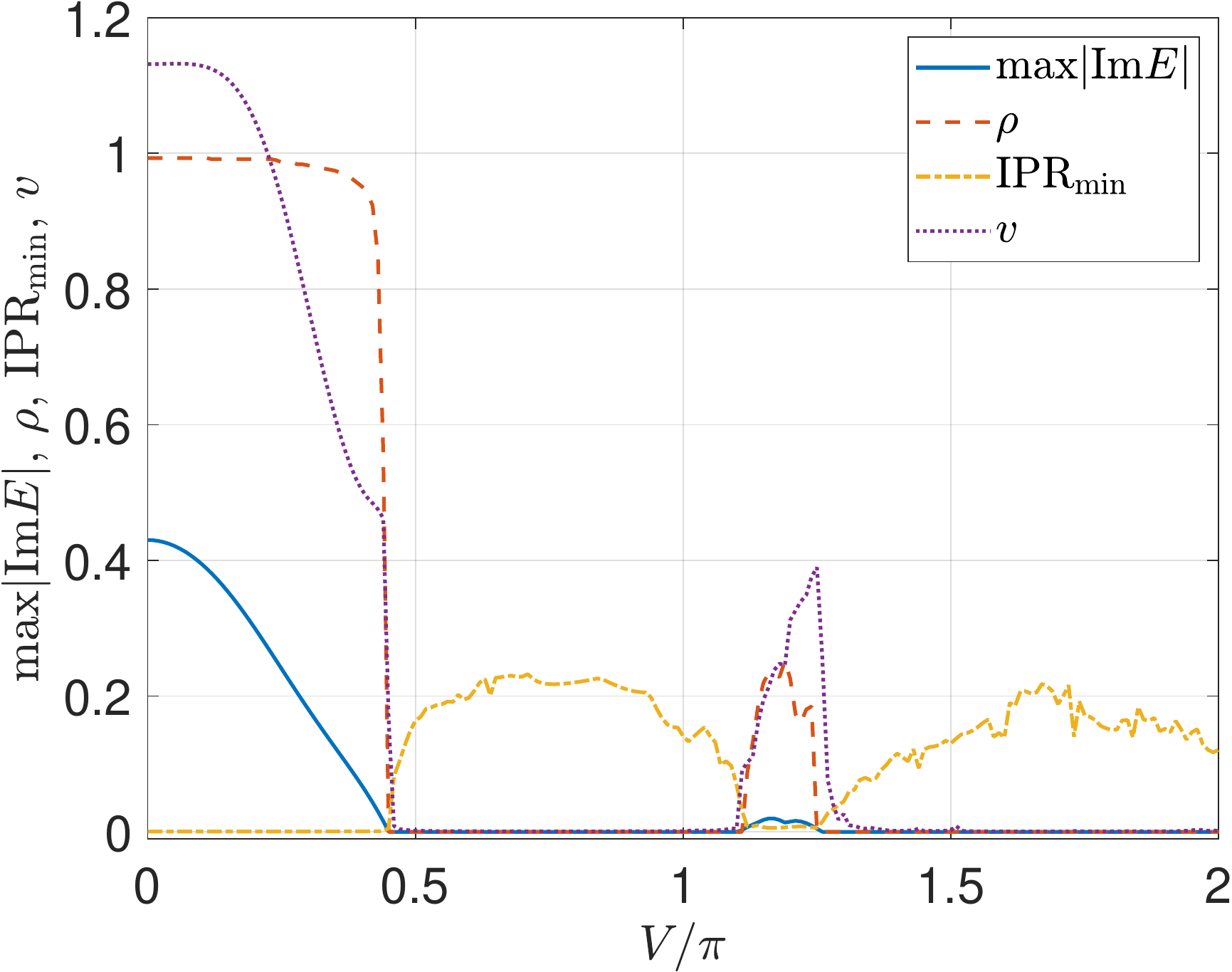}
		\par\end{centering}
	\caption{Maximal imaginary parts of quasienergy (blue solid line), DOSs with non-real quasienergies (red dashed line), minimum of IPRs (yellow dash-dotted line) and averaged spreading velocity a of wavepacket (purple dotted line) of the dimerized NRKHM versus the onsite potential amplitude $V$ under the PBC. Other system parameters are $(J,\gamma)=(\pi/6,0.8)$. The lattice size is $L=2584$. The initial state in the calculation of $v=v(t)$ is chosen to be $|\psi(0)\rangle=\sum_{n}\delta_{n0}|n\rangle$. The average in Eq.~(\ref{eq:VT}) is taken over $1000$ driving periods.}\label{fig:IPRmVtE-Dimer}
\end{figure}

In this appendix, we briefly discuss the transitions in the system described by the ${\hat H}_{\rm eff}$ in Eq.~(\ref{eq:Heff}). We truncate the series at the order $T^2$, i.e., to include all the terms explicitly shown in Eq.~(\ref{eq:Heff}). In Fig.~\ref{fig:IPRmVtE-Heff}, we present the properties of spectrum and IPR of the ${\hat H}_{\rm eff}$. The results demonstrate the presence of reentrant transitions between extended and critical mobility edge phases in the system at the level of ${\hat H}_{\rm eff}$. The is consistent with our argument in the main text that the reentrant localization transitions are originated from long-range couplings induced by the Floquet driving. However, no ${\cal PT}$ transitions and localized phases are observed for the considered domain of superlattice potential $V$. This is expected, as the driving frequency $\Omega=2\pi/T$ is equal to $\pi$ in our calculation, which is comparable with other parameters $J$ and $V$ of the system. In such a storng and near-resonant driving regime, higher order terms in ${\hat H}_{\rm eff}$ may also have important contributions, and one should not expect to capture the whole physics at the level of ${\hat H}_{\rm eff}$ by truncating the series in Eq.~(\ref{eq:Heff}) at a finite order. The problem is essentially non-perturbative. Moreover, the $2\pi$-periodicity of ${\hat U}$ in ${\mathsf V}_n$ is also lost at the level of ${\hat H}_{\rm eff}$, and the second part of our argument in Sec.~\ref{Dis} does not work in this case. Therefore, we emphasize that a full numerical treatment of the Floquet operator ${\hat U}$ is needed in order to capture the complete physics of quasienergy spectrum and localization transitions in the NRKHM (and also in general non-Hermitian Floquet quasicrystals) in strong and near-resonant driving regimes.

\section{Phase transitions in a dimerized NRKHM}\label{app2}

In this appendix, we briefly consider the ${\cal PT}$ and localization transitions in a dimerized variant of the NRKHM. Lattice dimerization may be introduced to the onsite potential or hopping amplitude. For simplicity, we consider a dimerized hopping model by replacing ${\hat K}$ in Eq.~(\ref{eq:K}) with
\begin{alignat}{1}
\hat{K}'& = J\sum_{n}(e^{\gamma}|2n-1\rangle\langle 2n|+e^{-\gamma}|2n\rangle\langle 2n-1|)\nonumber\\
& + J\sum_{n}(|2n\rangle\langle 2n+1|+{\rm H.c.}).\label{eq:Kprime}
\end{alignat}
Following the same quench protocol as given by Eq.~(\ref{eq:Ht}) in the main text, the resulting Floquet operator of the system reads ${\hat U}'=e^{-i{\hat V}T_2/\hbar}e^{-i{\hat K}'T_1/\hbar}.$ In Fig.~\ref{fig:IPRmVtE-Dimer}, we present the ${\cal PT}$ and localization transitions of this dimerized NRKHM versus the strength of onsite potential $V$ with $T_1=T_2=1$. The Floquet spectrum, IPR and wavepacket velocity are computed by the method sketched in Sec.~\ref{sec:Mod}. We find again multiple ${\cal PT}$ transitions, critical phases with mobility edges and reentrant localization transitions in this dimerized NRKHM. Nevertheless, the transition points and ranges of critical phase are different from the original model due to the effect of dimerized hopping. Meanwhile, lattice dimerization may also induce other interesting physics in non-Hermitian Floquet quasicrystals, such as richer phase patterns, more localization transitions and new topological properties. We leave a thorough investigation of this issue to potential future studies.


\end{document}